\documentclass[12pt]{article}
\usepackage[dvips]{color}
\usepackage{mathtext}
\usepackage[koi8-r]{inputenc}
\usepackage{graphicx}
\usepackage[figuresright]{rotating}
\usepackage{amssymb,eucal}
\usepackage{cite}
\date{}

\newcommand{\beq}{\begin{equation}}
\newcommand{\eeq}{\end{equation}}
\newcommand{\beqn}{\begin{eqnarray}}
\newcommand{\eeqn}{\end{eqnarray}}  
\begin{document}
\title{Performance of a CsI(Tl)  calorimeter in an experiment with stopped $K^+$'s }
%\author{Yury Kudenko\address{Institute for Nuclear Research RAS, \\
        %117312 Moscow, Russia} }
\author{Yu.G.~Kudenko\thanks{email:
          kudenko@wocup.inr.troitsk.ru}\\
	  ~\\
Institute for Nuclear Research RAS, 117312 Moscow, Russia \\
 For the KEK-E246 Collaboration}
\date{}
\maketitle 
\begin{abstract}
 The performance of the photon detector constructed for the search of 
  T-violation in the decay $K^+\to\pi^0\mu^+\nu$ 
 is presented. The specific features of this detector  consisting of 768 CsI(Tl) crystals with
 PIN photodiode  readout for high precision measurement of
T-odd correlations in decays of positive kaons are considered.  \\
%PACS: {\bf 29.40.Mc}\\
\end{abstract}
%\begin{keyword}
% CsI(Tl) calorimeter, PIN photodiode readout, positive kaons 
%\end{keyword}
\section{Introduction}
The main goal of  experiment E246 
 at the KEK 12~GeV proton synchrotron is a search for T--violating transverse
 muon  polarization ($P_T$)  in the decay 
 $K^+\to\pi^0\mu^+\nu$ ($K_{\mu3}$) with a sensitivity of 
 about $\Delta P_T \sim10^{-3}$.  The Standard Model predictions for $P_T$ is
 less than $10^{-7}$~\cite{bigi,valencia} and a contribution from the final 
 state interactions was found to be below the $10^{-5}$ level~\cite{fsi}. 
 So, this experiment is able to search for 
 additional or alternative sources of 
 CP-violation postulated in  various extensions to the Standard Model. 
 In particular  it is sensitive
 to the  CP-violation in the   multi Higgs doublet models, 
 the leptoquark models and supersymmetric models with/without 
 R--parity violation, where $P_T$ can reach the measurable level of
 $10^{-4}-10^{-2}$~\cite{weinberg}. 
 
\begin{figure}[htbp]
\centering\includegraphics[width=16cm,angle=0]{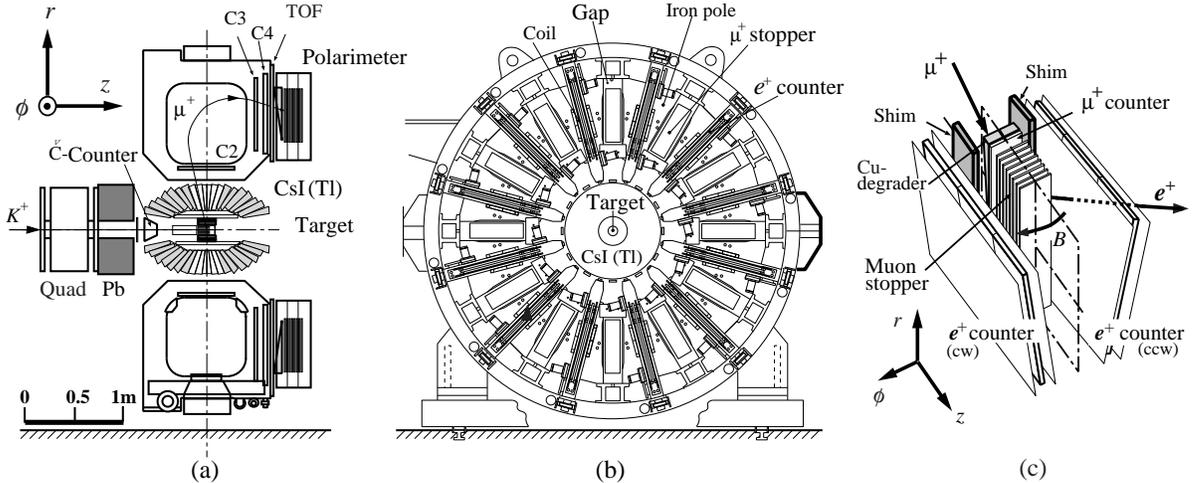}
\caption{Experimental setup; (a) side view, (b) front view, and 
(c) one sector of the polarimeter.}
\label{fig:setup}
\end{figure}

 The E246 detector 
 was commissioned in 1996 and accumulated data for about 6 years. First results 
 with a new limit on $P_T$  were published in~\cite{prl99}. 
 In this experiment, the $K_{\mu3}$ decay is identified by detecting the
 $\pi^0$ as well as the $\mu^+$ from the decay. The layout of the E246 
 set--up is shown in Fig.~\ref{fig:setup}.
%----------------------  
%----------------------
 A 660~MeV/c 
 kaon beam is slowed down in a BeO degrader and stopped in an active target,
 made of 256 $5\times5$~mm$^{2}$ scintillating fibers.     
 The energy and direction of the $\pi^{0}$  from the $K_{\mu3}$ decay are 
 measured by a segmented CsI(Tl) photon detector installed in the central region of the toroidal magnet. A muon from the 
 $K_{\mu3}$ decay at rest is momentum--analyzed by one of the 12 magnet gaps
 of the superconducting toroidal spectrometer  using tracking performed by 
 the stopping target,  a scintillating ring 
 hodoscope~\cite{ring_hodoscope} surrounding 
 the target,  and three MWPCs. The muon exiting the spectrometer   is stopped 
 in a polarimeter 
 in which the decay positron asymmetry $A_T$ is measured in order to obtain 
  $P_T$. The polarimeter consists of 12 azimuthally arranged Al stoppers, 
  aligned 
 with the magnet gaps, with a scintillator counter system located between the 
 stoppers.
 A positron from the decay of $\mu^+\rightarrow e^{+}\nu\nu$ tends to be 
 emitted
 in the same direction as the muon spin, and is detected by these plastic 
 scintillator counters.
 In this detector $A_T$ appears as a difference in the counting
 rate between clockwise ($cw$) and counter-clockwise ($ccw$) emitted positrons. 
 By summing the $cw$ ($N_{cw}$) and $ccw$ ($N_{ccw}$) positron counts  over all 
 12 sectors, $P_T$ is derived from
\begin{equation}
 P_T = \frac{1}{\alpha f} \cdot \frac{N_{cw} -  N_{ccw}}
 { N_{cw} +  N_{ccw}},
 \end{equation}
where  $\alpha$ is the analyzing power of the polarimeter and  $f$ is 
 an attenuation factor which reflects the fact that the  decay plane of the 
 $K_{\mu3}$ event determined from the $\mu^+$ and $\pi^0$ momenta is not 
 parallel to the median plane of the gap in which the muon momentum is measured.

The  photon detector plays a crucial role in this experiment for suppression of
systematic errors. Using complete reconstruction of the $K_{\mu3}$ kinematics 
  we separate  these events into
 two classes: events with the pion moving along the kaon beam (forward 
 direction)
 and those where the pion moves in the backward direction. 
 The sign of 
 P$_T$ for the
 forward-going $\pi^0$ events is opposite to that for backward-going 
 $\pi^0$ events.  
 This allows us to apply a double ratio
 between the forward- and backward-going $\pi^0$ events  cancelling out most   
 instrumental sources of false polarization, which are likely
to be independent of the $\pi^0$ direction.
 This method is an unique feature of E246 experiment, and  is a key factor
  for effective suppression of  most systematic errors. 
  
\section{Overall  description of the calorimeter}
\begin{figure}[htbp]
\centering\includegraphics[width=12cm,angle=0]{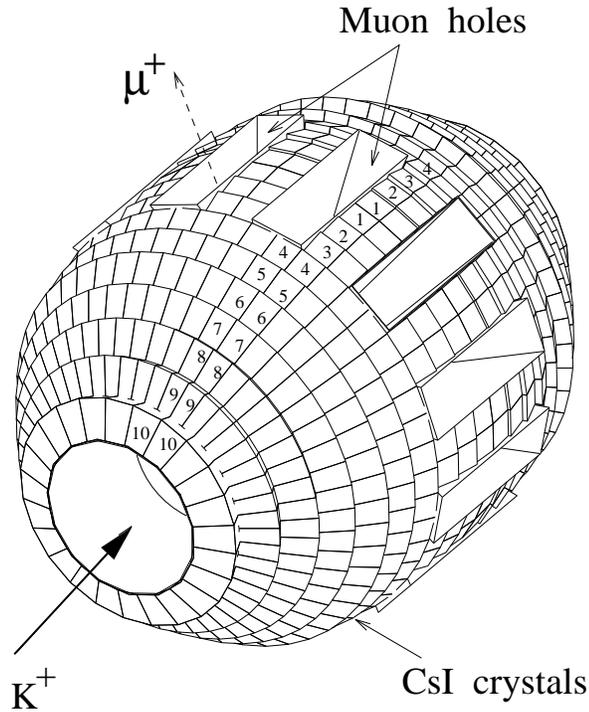}
\caption{The photon detector. }
\label{fig:barrel}
\end{figure}

The photon detector
 consists of 768  CsI(Tl) crystals each with a PIN photodiode  readout.  They form
 a barrel structure that covers  about 75\% of $4\pi$ with 12 symmetrically 
 spaced 
 holes  for $\mu^{+}$s going into the magnetic spectrometer 
 (Fig.~\ref{fig:barrel}).
%-------------- 
%---------------------- 
 
 The detector is capable  detecting photons emerging from the target in the
 polar angle range from 15$^{\circ}$ to 165$^{\circ}$ and in 2$\pi$ of the azimuthal
 angle except for the muon holes. The optically 
 isolated CsI modules point towards the center of the active target. An individual
 crystal covers 7.5$^{\circ}$ in both polar and azimuthal angles, except for   48 crystals
 near to the beam axis, where the azimuthal angle was doubled to 15$^{\circ}$, 
 and 
 the muon hole region. The crystal length of 250~mm (13.5~r.l.) was chosen 
 in order to leave   about 4 cm  of space available  radially to mount  a 
 PIN-diode and  preamplifier. The overall gain stability of the calorimeter 
 is monitored by a Xe--lamp light pulser 
 system with optical fibers which distribute the light from
 the lamp to the crystals.
 
\begin{figure}[htbp]
\centering\includegraphics[width=15cm,angle=0]{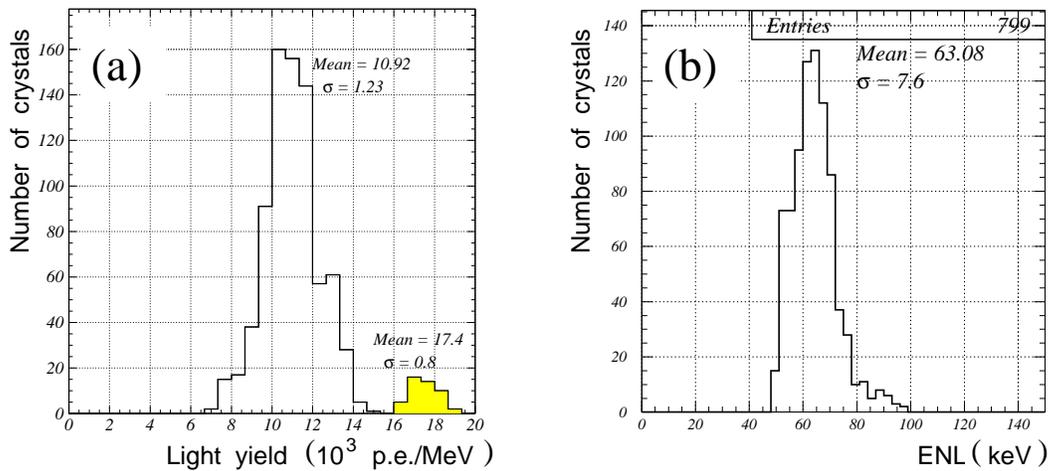}
\caption{(a) Distribution of the light yield. The shadowed region
 of 16000--20000 p.e./MeV corresponds to the modules which have a PD
 with $28\times28$~mm$^{2}$ sensitive area. (b) Distribution of 
 the equivalent noise level. }
\label{fig:enl}
\end{figure}

 To obtain uniform high light yield along the crystal, the crystal surfaces were treated to grade the reflectivity, and a Millipore white paper 
 was chosen as a reflector (see for details Ref.~\cite{csi_pte}). 
 As a result, for almost all modules the light yield of more than 
 8000 p.e./MeV  and
 the average equivalent noise level (ENL) of about 63 keV were obtained, 
 as shown in  Fig.~\ref{fig:enl}.
 %----------
%--------- 
 The whole detector demonstrated a low 
 level of coherent noise of about 11 keV ($\sigma$) per
 module, which has only a small effect on the energy resolution 
 and signal-to-noise ratio of the detector.
 These parameters are the best among the 
 large CsI(Tl)  calorimeters which have been built in recent time~\cite{cleo}. 
 
\begin{figure}[htbp]
\centering\includegraphics[width=13cm,angle=0]{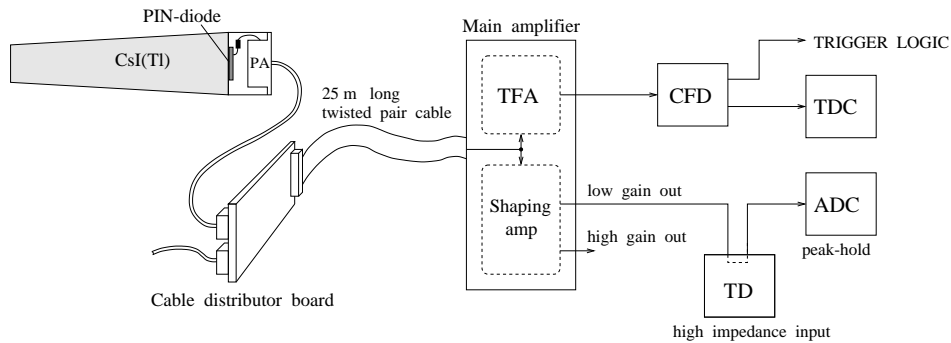}
\caption{Electronics of a CsI module. }
\label{fig:electr}
\end{figure}

 To efficiently use the high light yield and low ENL of 
 individual crystals,  the 
 electronics  for the calorimeter was  specially designed. The high counting
 rate environment in the calorimeter due to  pion contamination in the beam 
 was also taken into account in the design. The calorimeter electronics
  is schematically shown
 in Fig.~\ref{fig:electr}
 %-----------
%--------- 
 and described in detail in Ref.~\cite{electronics}. 
 The main amplifier incorporates a shaping amplifier with fast restoration 
 of the base level and a timing filter amplifier which forms 
 a timing signal  for a subsequent discrimination by a constant fraction 
 discriminator (CFD).  The CsI timing signals from all CFD outputs are 
 used in the second level trigger in coincidence with the first level fast 
 trigger. 
 
 The detector stability is monitored by a Xe--lamp system. During the nearly 
 6 year period of operation none of the 768 photodiodes and only two preamplifiers  failed.  A  light yield degradation of the CsI modules of about 
 12\% was observed over this time.

 \section{Performance}
 \subsection{In-beam calibration}
\begin{figure}[htbp]
\centering\includegraphics[width=14cm,angle=0]{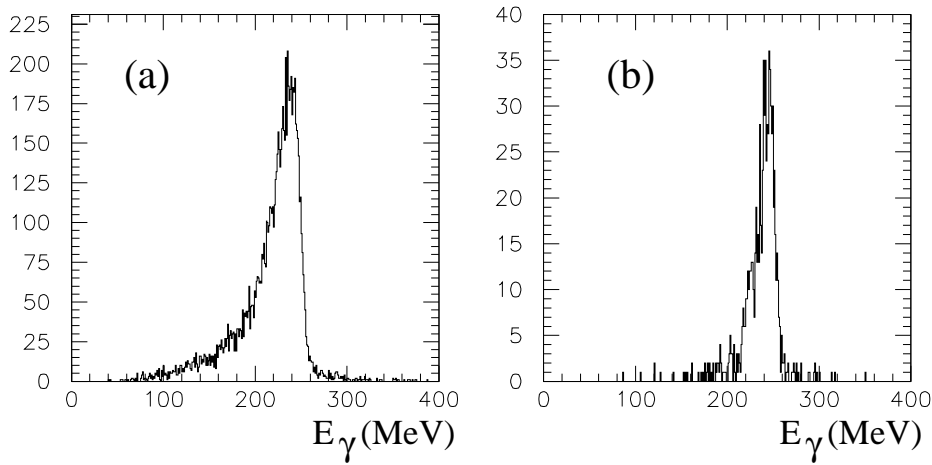}
\caption{Energy sum of two $5\times 5$ photon clusters from the 
$K_{\pi2}$ decay. (a) All  events with peak value $E_{\gamma\gamma} = 235.7$ 
MeV, $\sigma = 5.2\%$.  The low energy tail is due to the shower leakage 
into muon and beam holes. (b) Photons  are detected away from muon holes. 
$E_{\gamma\gamma} = 242.5$ 
MeV, $\sigma = 4.1\%$.
   }
\label{fig:kpi2}
\end{figure}

 The in-beam calibration of the detector was carried out using muons from 
 the dominant decay mode  $K^+\to \mu^+\nu$ of stopped kaons. 
 These muons have a kinetic energy 
 of 152.4~MeV and are stopped in the
 CsI crystals. The  calibration runs, which were carried out 
 every beam cycle, used a reduced beam intensity and a special trigger. 
 It included the kaon identification with the Cherenkov detector, kaon stop in the target, 
 and the presence of a signal in the CsI. In--flight kaon 
 decays were avoided by introducing a 15--ns time delay.   
 The events in which no energy was detected in the neighboring
 crystals were selected to obtain the individual gains of CsI modules. The
 kaon decay vertex  was reconstructed from $\mu^+$ tracking  and muon energy 
 losses  in the target were
 corrected using the target fiber ADC  information~\cite{csi_calibr}. 
 Using individual CsI calibration coefficients the decay $K^+\to\pi^+\pi^0$
 is reconstructed.
 Fig.~\ref{fig:kpi2}
 %---------
%--------- 
 shows the spectrum of the energy sum of two photon 
 clusters from the  decay
 $\pi^0\to\gamma\gamma$. The peak value 235.7 MeV corresponds to about 96\% of 
 the $E_{\pi^{0}}$=245.6 MeV, that is consistent with the result of Monte 
 Carlo calculation for the cluster size of
 5$\times$5 crystals. The low energy tail is due to the shower leakage
 into the muon  and beam holes, and through the rear sides of the crystals. 
 The  resolution $\sigma_{E}/E$ obtained for the  energy sum of two photons 
 from  $K_{\pi2}$ decay is 5.2\%.  The  energy 
 resolution is better for clusters away 
 from muon holes, as seen from Fig.~\ref{fig:kpi2}(b).  The invariant mass
 resolution of the $\pi^0$ was  6.7\% for whole detector and 
 5.6\% for events with both photons detected away from muon holes.
 \subsection{Time resolution} 
\begin{figure}[htbp]
\centering\includegraphics[width=11cm,angle=0]{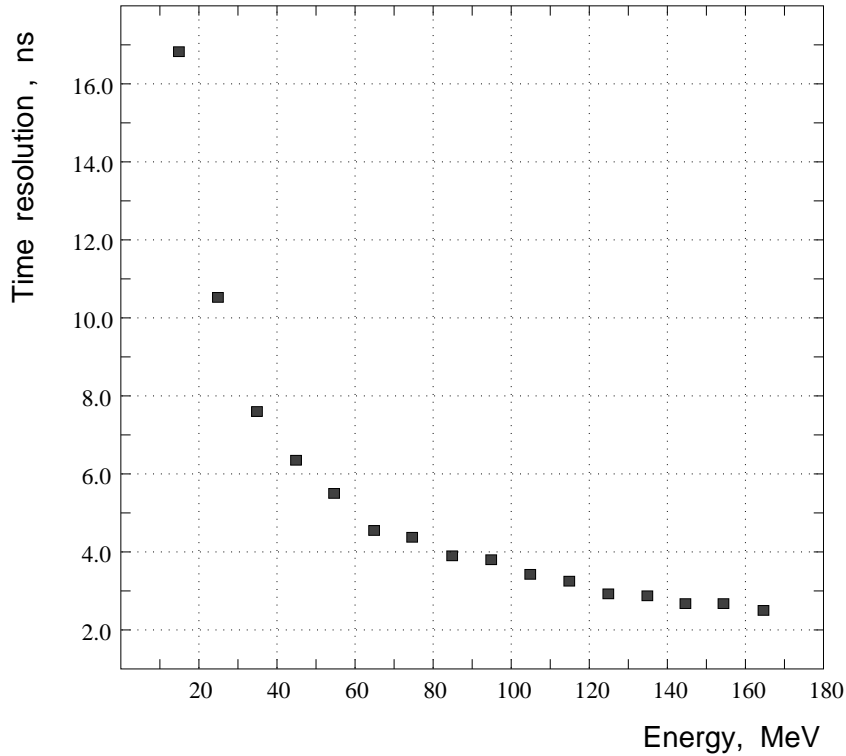}
\caption{Time resolution as a function  of the photon energy.  }
\label{fig:time}
\end{figure}

 Good time resolution is an important factor for reliable 
 identification of photons and neutral pions from the decays of $K^+$'s and
 for background suppression.
 The time resolution was determined in physics runs by
 measuring time difference between a muon signal from $K_{\mu3}$ decay
  produced by
 one of the 12  fiducial plastic counters surrounding  the target, 
 and a CFD signal from a CsI module.  Fig.~\ref{fig:time}
%----------
%-----------  
 shows energy 
 dependence
  of the obtained time resolution ($\sigma$) of the CsI crystals which are 
  the centers of the gamma clusters. 
 Time resolution of 3.5~ns  at 100~MeV deteriorates to 16.8~ns in the
 low energy range of 10--20~MeV.  For the $K_{\mu3}$ photon energy range of 
 10-250 MeV the resolution $\sigma = 3.8$ nsec was obtained. 
 To exploit the timing performance,  all the CsI CFD outputs were 
 added to the trigger logic as the OR strobe.  A 150 ns discriminator 
 pulse width was sufficient to accept nearly all good events, 
while reducing the recorded trigger rate by 
 a factor of 2.
\subsection{Counting rate}
\begin{figure}[htbp]
\centering\includegraphics[width=12cm,angle=0]{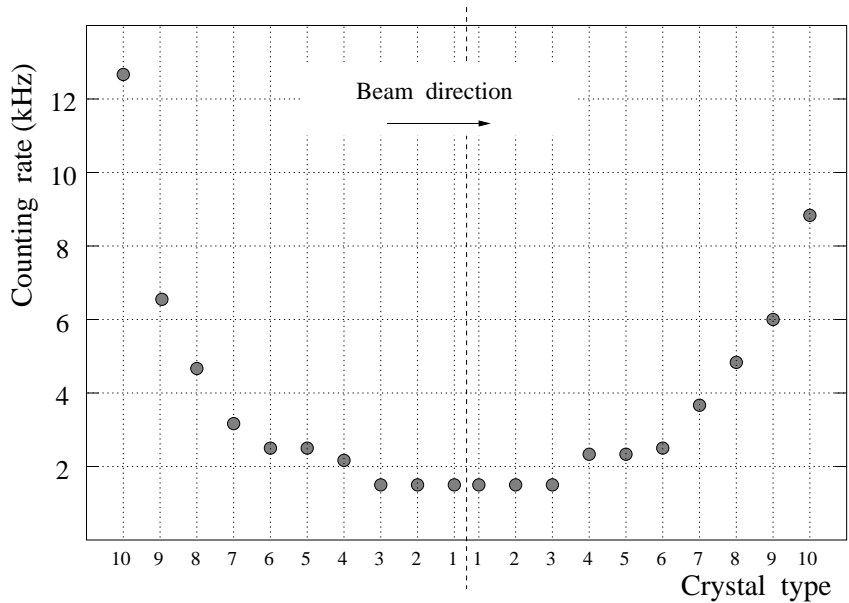}
\caption{Countig rate in individual crystals. Position of each type of 
crystal in the calorimeter is shown in Fig.~\ref{fig:barrel}. }
\label{fig:rate}
\end{figure}

The slow decay time of CsI(Tl) crystals can lead to losses of detected events in
a high intensity beam due to pile-up of the CsI pulses. The large
contamination of pions in the beam ($\pi/K\geq 6$)  and an asymmeterical pion 
halo were the main sources which dominated the counting rate in the calorimeter.
The counting rate of a crystal depends on its position relative to the beam axis 
and was dramatically increased for the crystals closest to the beam, as seen 
in Fig.~\ref{fig:rate}.
%--------- 
%--------- 
Taking into account the distribution of the events
over the solid angle, the average counting rate per ctystal in the experiment 
was determined to be about 8 kHz. The tests of baseline stabilty at this 
counting rate using Xe--lamp events showed  that about 5\% of events were 
lost in physics runs due to pile up.  
\subsection{Detector alignment}
\begin{figure}[htb]
\centering\includegraphics[width=12cm,angle=0]{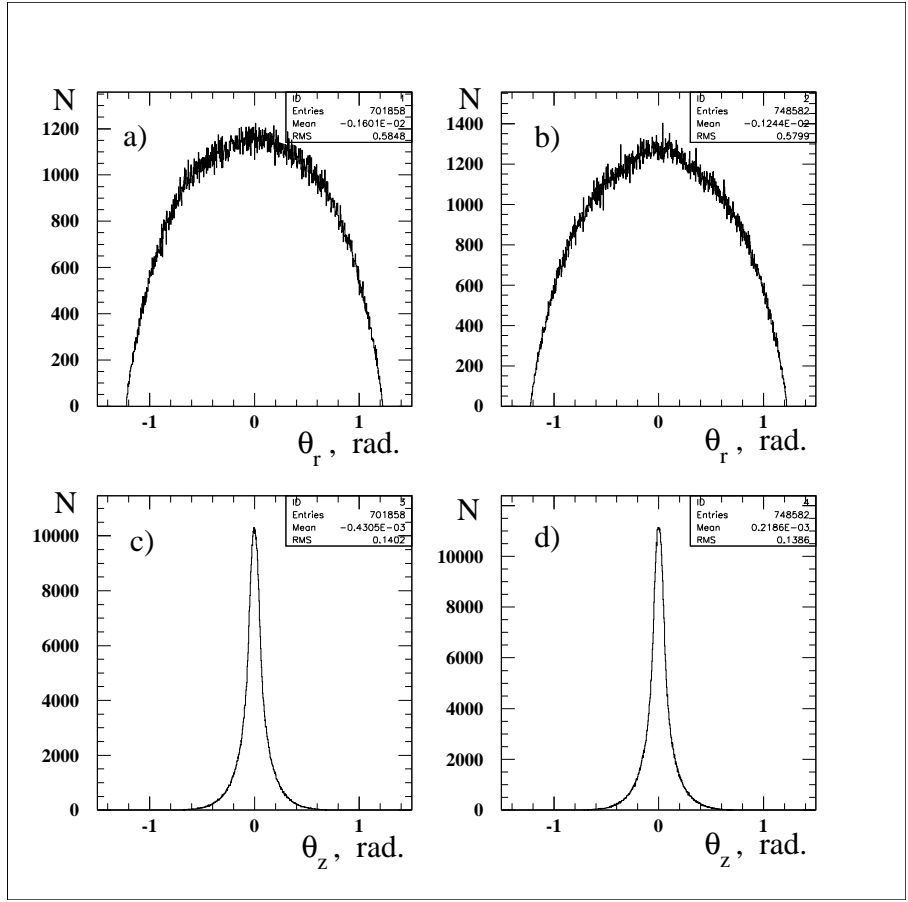}
\caption{The $K_{\mu3}$ decay plane rotation around r and z axes. (a) and (c) 
$K_{\mu3}$ events with forward-going pions;  (b) and (d) $K_{\mu3}$ events 
with backward-going pions.}
\label{fig:decay_rotation}
\end{figure}

The  limited accuracy of the detector installation can make  an
 asymmetry in
the angular distribution between the median plane of the polarimeter and 
$K_{\mu3}$ decay plane, the normal vector $\vec n$ to  which is defined as
\begin{equation}
\vec n = \frac{\vec{p}_{\pi^{0}}\times\vec{p}_{\mu^{+}} }
{\mid \vec{p}_{\pi^{0}}\times\vec{p}_{\mu^{+}}\mid }.
\end{equation}
Here $\vec{p}_{\pi^{0}}$ and $\vec{p}_{\mu^{+}}$ are the momenta of
the pion and muon, respectively. This angular asymmetry can give rise to
contamination of the in-plane polarization and can generate a spurious 
T-odd effect due to the opposite direction of this 
polarization for the forward- and backward-going pions. According to the Monte 
Carlo simulation, the
residual value of the muon in-plane polarization in the polarimeter is  0.5--0.7 
 for both directions of the  neutral pions.
The  measured angular distributions of $K_{\mu3}$ decay plane 
rotations around the beam direction (z-axis) and radial direction 
(r-axis, perpendicular to the beam and lying in the median plane of a magnet 
sector, see Fig.~\ref{fig:setup}(c))  are presented in 
Fig.~\ref{fig:decay_rotation}.
%-----------
%--------- 
As one can see, the detector provides good
angular symmetry, and possible spurious polarization caused by detector 
misalignment is $\leq 3.7\times 10^{-4}$, i.e. much less than the expected
statistical error.    
\subsection{Double ratio} 
\begin{figure}[htbp]
\centering\includegraphics[width=6cm,angle=0]{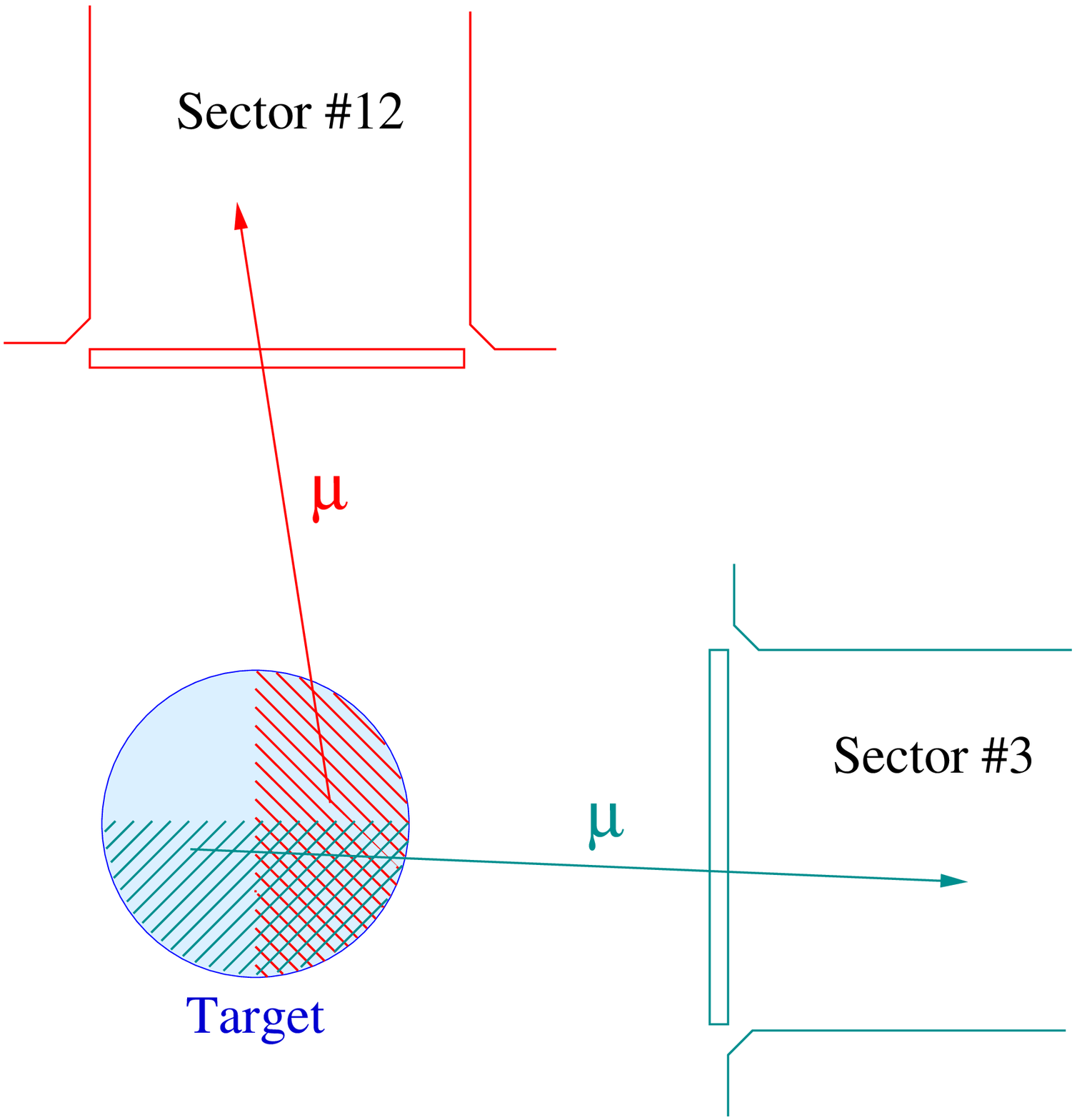}
\caption{Schematic view of the events selection for study of the 
systematics compensation by double ratio}
\label{fig:half_target}
\end{figure}

\begin{figure}[htbp]
\centering\includegraphics[width=11cm,angle=0]{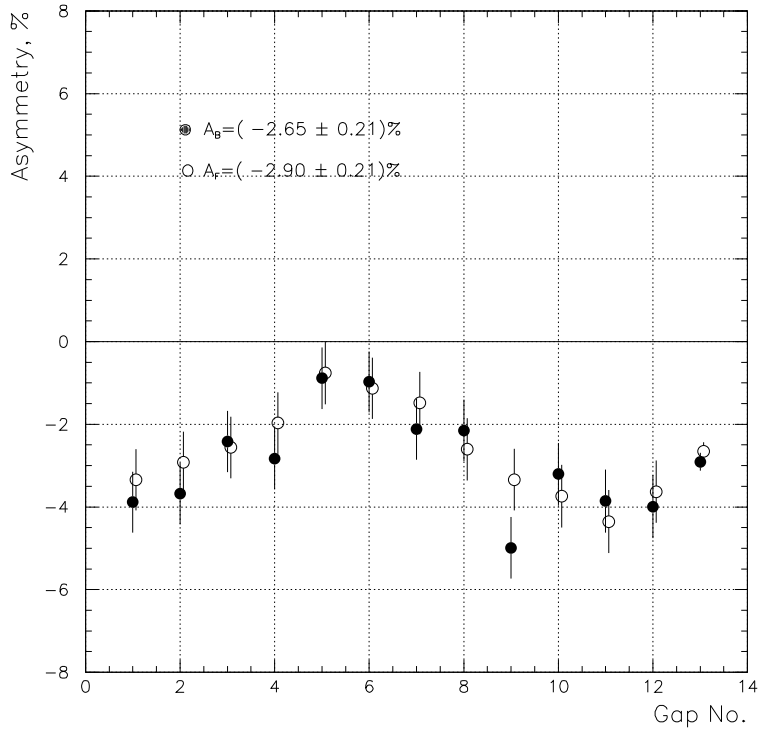}
\caption{The illustration of the suppression of systematic errors using  double
ratio. The open (solid) circles represents $K_{\mu3}$ events with forward
(backward)-going  pions selected as shown in Fig.~\ref{fig:half_target}. 
The values shown for  Gap No.13 are sums of forward and 
backward asymmetries over all polarimeter sectors.
}
\label{fig:double}
\end{figure}

To measure the power of the double ratio, the compensation of a large 
imitation asymmetry in the experimental kaon 
stopping distribution was studied for the good forward and 
backward $K_{\mu3}$ events. The events accumulated in 1998 running cycle 
were selected, in the right-hand part of the target relative to the
median plane in each magnet sector looking upstream the beam, as shown in 
Fig.~\ref{fig:half_target}.
%----------
%--------- 
As a result, 
in each gap a spurious asymmetry of 2--6\% appears, as seen in 
Fig.~\ref{fig:double},
%-----------
%----------- 
 and total asymmetry is $A_F = -2.90\pm 0.21\%$ and 
$A_B = -2.65\pm 0.21\%$ for  forward and backward events, respectively. So,   
the double ratio 
\begin{equation}
(A_F + A_B)/(A_F - A_B) \simeq 22
\end{equation}
provides a large 
factor  for cancellation of systematic errors in the asymmetry measurement.
For different   data sets this factor was determined to be in the range of 15-24.
\section{Conclusion}
The CsI(Tl) photon detector with PIN photodiode readout is being successfully 
used  in the 
E246 experiment at KEK. The overall performance of the calorimeter
 was not degraded significantly after being in operation for 6 years.
 The CsI calorimeter provides good reconstruction of the $\pi^0$ parameters 
 and seems to be a key instrument for reduction of systematic 
 uncertanties in the polarization measurement down to the level of $10^{-3}$ 
 that completely fulfills the requirements of the experiment.

 %----------------------------------------------------------------------------

%-------------------------------------------------------------------------------------  


\begin{thebibliography}{99}
\bibitem{bigi}I.I.~Bigi and A.I.~Sanda, CP violation, Cambridge University
Press, 2000.
\bibitem{valencia}E.~Golowich and G.~Valencia, Phys. Rev., {\bf D40} (1989) 
112. 
\bibitem{fsi}A.R.~Zhitnitskii, Yad. Fiz. {\bf 31} (1980) 1024.\\
V.P.Efrosinin et. al., Phys. Lett. {\bf B493} (2000) 293, hep-ph/008199.
\bibitem{weinberg}S.~Weinberg, Phys. Rev. Lett. {\bf 37} (1976) 657.\\
G.~B\'{e}langer and C.Q.~Geng, Phys. Rev. {\bf D44} (1991) 2789.\\
G.-H.~Wu and J.N.~Ng, Phys. Lett. {\bf B392} (1997) 93, hep-ph/9609314.\\
M.~Fabbrichesi and F.~Vissani, Phys. Rev. {\bf D55} (1997) 5334, 
hep-ph/9611237.
\bibitem{ring_hodoscope}A.P.~Ivashkin et al.,  Nucl. Inst. Meth. {\bf A394} 
(1997) 321.
\bibitem{prl99}M.~Abe et al., Phys. Rev. Lett. {\bf 83} (1999) 4253.
\bibitem{csi_pte}M.P.~Grigorev et. al., Inst. Exp. Tech. {\bf 39} 
(1996) 164 [Prib. Tekhn. Eksp. {\bf N2} (1996) 18].
\bibitem{csi_nim}D.V.~Dementyev et al., Nucl. Instr. Meth. {\bf A440} (2000) 
151. 
\bibitem{cleo} Y.~Kubota et al., Nucl. Instr.  Meth. {\bf A320} (1992) 66.\\
E.~Aker et al., Nucl. Instr.  Meth. {\bf A321} (1992) 69.\\
B.~Shwartz, ``Crystal calorimeters'', talk at this Conference, 
http://www.inp.nsk.su/events/confs/instr2002/day5.shtml; 
K.~Miyabayashi, ``BELLE electromagnetic calorimeter'', ibid;
B.Lewandowski, ``Design and performance of the BaBar EM calorimeter'', 
ibid.
\bibitem{electronics} Yu.G.~Kudenko, O.V.~Mineev, J.~Imazato,
Nucl. Instr. Meth. {\bf A411} (1998) 437.
\bibitem{csi_calibr}M.M.~Khabibullin et al., Instr. Exp. Tech. 
{\bf 43} (2000) 589 [Prib. Tekhn. Eksp. {\bf N5} (2000) 9].
 
\end{thebibliography}
\end{document}